\documentclass[acmsmall]{acmart}

\AtBeginDocument{%
  \providecommand\BibTeX{{%
    \normalfont B\kern-0.5em{\scshape i\kern-0.25em b}\kern-0.8em\TeX}}}

\setcopyright{acmcopyright}
\copyrightyear{2023}
\acmYear{2023}
\acmDOI{XXXXXXX.XXXXXXX}

\acmConference[CSCW '23]{The 2023 ACM SIGCHI Conference on Computer-Supported Cooperative Work & Social Computing}{Oct 13--18,
  2023}{Minneapolis, MN}
\acmPrice{15.00}
\acmISBN{978-1-4503-XXXX-X/18/06}

\begin{document}

\title{Twitter has a Binary Privacy Setting, are Users Aware of How It Works?}


\author{Dilara Keküllüoğlu} 
\affiliation{%
  \institution{University of Edinburgh}
  \department{School of Informatics}
  \city{Edinburgh}
  \country{UK}}
\email{d.kekulluoglu@ed.ac.uk}

\author{Kami Vaniea}
\affiliation{%
  \institution{University of Edinburgh}
  \department{School of Informatics}
  \city{Edinburgh}
  \country{UK}}
\email{kvaniea@inf.ed.ac.uk}

\author{Maria K. Wolters}
\affiliation{%
  \institution{University of Edinburgh}
  \department{School of Informatics}
  \city{Edinburgh}
  \country{UK}}
\email{Maria.Wolters@ed.ac.uk}

\author{Walid Magdy} 
\affiliation{%
  \institution{University of Edinburgh}
  \department{School of Informatics}
  \city{Edinburgh}
  \country{UK}}
\email{wmagdy@inf.ed.ac.uk}
\renewcommand{\shortauthors}{Keküllüoğlu, Vaniea, Wolters, and Magdy}

\newcommand{\change}[1]{#1}
\newcommand{\minorchange}[1]{#1}
\begin{abstract}
 Twitter accounts are public by default, but Twitter gives the option to create protected accounts, where only approved followers can see their tweets. The publicly visible information changes based on the account type and the visibility of tweets also depends solely on the poster's account type which can cause unintended disclosures especially when users interact. We surveyed 336 Twitter users to understand users’ awareness of \change{account} information \change{visibility}, as well as the tweet visibility when users interact. We find that our participants are aware of the visibility of their profile information and individual tweets. However, the visibility of followed topics, lists, and interactions with protected accounts is confusing. Only 31\% of the participants were aware that a reply by a public account to a protected account’s tweet would be publicly visible. Surprisingly, having a protected account does not result in a better understanding of the \change{account} information or tweet visibility.
\end{abstract}



\maketitle

\section{Introduction}

Twitter is a popular micro-blogging platform that provides users with limited low fidelity privacy controls which have the potential to make Twitter's privacy easier for users to understand. However, the functional reality is that the way controls combine across time and between accounts may make them deceptively confusing. 
For example, when a user changes their account visibility on Twitter it changes not only the visibility of future tweets but also past ones. Similarly, when a user with a public account replies to the tweet of a private account that they follow, the response is public. Existing research shows that users share potentially private information on Twitter~\cite{kekulluoglu2020analysing} and underestimate the reach of their posts~\cite{bernstein2013quantifying}. Most of these prior works might assume that Twitter users have a good understanding of privacy controls thanks to the control's relative simplicity compared to the more detailed settings on platforms like Facebook. In this work we test this assumption by exploring how well Twitter users understand the impacts of the privacy settings.

Twitter frames itself as a place where users can share posts and connect with others. Twitter users can share various information about themselves publicly through their profile by including such information as a profile picture, bio, location, and so on regardless of their account type. Users can also have \emph{conversations} with other users by replying back and forth to tweets, creating a \emph{thread}. As well as engage with others by \emph{mentioning} them, \emph{liking}, \emph{retweeting}, or \emph{quote tweeting} their tweets. 
Twitter offers a binary privacy configuration where accounts are either public (default) or protected, which affects the visibility of account activities. Tweets from public accounts can be seen by anyone on the Internet, while tweets by protected accounts can only be seen by the account's approved followers. \change{Some users also utilize privacy settings of Twitter dynamically where they switch their settings between public and protected to enjoy the full functionalities of the platform while protecting their privacy~\cite{kekulluoglu2022understanding}. }However, only 4\% of Twitter users have protected accounts~\cite{liu2014tweets}, so the visibility \change{of information} associated with them \change{is} not necessarily well understood.

The binary privacy options are fairly easy to understand at the surface level so it might be expected that most users understand the implications of the public/protected configuration. However, inadvertent disclosure still happens on Twitter~\cite{kekulluoglu2020analysing}. \change{One of the reasons for these unwanted data leaks on social media is the lack of user understanding around privacy settings~\cite{madejski2011failure}. Interactions further complicate the situation, especially when the interacting parties have different privacy settings, creating a ''boundary turbulence``~\cite{petronio2015communication}. These interactions can leak information in a way that cannot be easily controlled by the information owner~\cite{jurgens2017writer,kekulluoglu2020analysing,aldayel2019your} which can have unforeseen consequences such as insurance premium increases~\cite{insurancesocialmedia}, identity fraud by data chaining~\cite{fraudsocialmedia}, or even losing disability benefits~\cite{insurancedisability}.}   
The simplicity of the configuration options may be leading users to a false sense of confidence where they believe that a post or account information will only be seen by a restricted set of followers, when that is not actually the case\change{~\cite{kekulluoglu2022authentication}}. Without this strong understanding it becomes difficult for users to enact their privacy intentions on Twitter.

\change{Most of the prior studies on privacy settings and visibility are conducted on Facebook where the privacy settings are granular and complex. However, there is limited research on platforms have limited low fidelity privacy controls, such as Twitter, where the privacy settings apply to the whole account, once changed the settings will affect future tweets as well as the past ones.} In this work, we want to gauge the users' awareness regarding the visibility of user information on Twitter, as well as the tweet visibility especially when users interact with each other. Specifically, we investigate the following research questions:
\begin{description}
    
    \item[RQ1] How well do Twitter users understand the visibility of user information and tweets in relation to different privacy settings? 

   \item[RQ2] What factors (e.g. social media experience, frequency of use, interactions with other accounts) contribute to users' awareness of the visibility of information and tweets? In particular, what is the role of the user's account type on their awareness \change{(i.e. public, protected, or switching)}?
   
\end{description}



To answer our research questions, we conducted a user survey with 336 participants who have Twitter accounts with a range of privacy settings, including participants who only use public account, protected account, and some switching between the two settings. Our findings show that the participants are mostly aware of who can see tweets of users when they are tweeting by themselves, i.e. not interacting with other users. They also mostly understand what account information is publicly visible with the exception of topics and lists. Interactions between public and protected accounts was more confusing with only 40\% of these questions answered correctly. Surprisingly, the normal audience (public, protected, switching) of the participants did not have any significant impact on their knowledge of the platform functionality around privacy settings. However, the frequency of replying to protected accounts, Twitter usage, and being able to easily see that they are interacting with a protected account have an impact on our participants' Twitter privacy functionality understanding. Our contribution\change{s include (1) measuring the levels of understanding for individual information visibility, as well as the interaction visibility on Twitter, (2) investigating the factors contributing to this understanding, and (3)} comparing \change{the functionality and privacy} awareness of users with different privacy settings, including the ones who \change{switch between settings} frequently. Our findings suggest that the design of Twitter UI might be sub-optimal, especially when it comes to dealing with protected accounts, where users are not fully aware of visibility of some of their activities on the platform. We list possible privacy violations that could happen in the platform and suggest design implications.

\section{Related Work}


Users bond  with each other over social networks gaining social capital while also wanting to protect their privacy~\cite{vitak2015balancing}. Aside from the sharing their daily lives and opinions, seeking support~\cite{dym2018vulnerable,andalibi2016understanding}, building reputation~\cite{syn2015social}, and utilizing it for professional contexts~\cite{mahrt2014twitter,marwick2011tweet} are some ways people use social media. Social networks provide privacy settings to protect users' privacy, giving a sense of control that may lead to more self-disclosure~\cite{stutzman2011factors,liang2017privacy}. Unfortunately, the range and scope of shared content can be unclear. Bernstein et al.~\cite{bernstein2013quantifying} analyzed 220k Facebook posts and found that users underestimate the true size of the audience by a factor of four. Such audience estimations are even harder on platforms like Twitter, where posts are mostly public and interactions by others increases the post's visibility.

People tune privacy configurations~\cite{marwick2014networked, madejski2011failure,kekulluoglu2022understanding}, censor themselves~\cite{marwick2011tweet,das2013self,sleeper2013post}, limit connections~\cite{johnson2012facebook}, create multiple accounts for different purposes~\cite{vitak2015balancing,stutzman2012boundary}, as well as stop using the platform~\cite{baumer2013limiting, grandhi2019stay,lampe2013users} to protect their privacy online. While somewhat effective, these strategies will not provide absolute privacy protection, especially against people within the user's social circle. The networked nature of the online social media causes protecting privacy to be a collective work~\cite{boyd2012networked,marwick2014networked}. Connections over social networks want to protect the user's privacy, however, disagreements over what is private can lead to unwanted disclosures~\cite{jung2016imagined}. Friends can share/re-share posts about users, disclosing information that is not easily controlled by the user~\cite{jurgens2017writer,jain2013call,meeder2010rt}. Sensitive information can be inferred from public replies to a post even when the user actively chose to protect their privacy~\cite{kekulluoglu2020analysing}. 

It is also not trivial for users to configure privacy settings to match their intended audience~\cite{liu2011analyzing,madejski2011failure,hoyle2017viewing}. Prior work has looked at context collapse where users' different social circles, such as friends, family, and colleagues, are all on the same social network~\cite{nissenbaum2009privacy,marwick2011tweet,vitak2012impact} making it hard for users to control the flow of information to different social circles. The temporal persistence of social media posts further complicates the situation~\cite{brandtzaeg2018time,huang2020you}. Intended audiences can also change overtime, e.g. a user regretting their decision to post~\cite{sleeper2013read}. Unclear permissions or lack of understanding regarding the social media permissions by the users can cause privacy violations.  Madejski et al.~\cite{madejski2011failure} measured the discrepancies between privacy settings and the intentions of the users on Facebook. They found that every participant reported at least one sharing violation. Hoyle et al.~\cite{hoyle2017viewing} studied LinkedIn users' understanding of ``viewed by'' feature including what controls they had available around it. They found that most people do not understand how the permission works.  These violations can result in unintended consequences, users may lose their job~\cite{wang2011regretted}, or insurance companies can increase their premiums~\cite{mao2011loose}.

Effective utilization of these privacy settings may be achieved by increasing general internet skills~\cite{buchi2017caring} and privacy literacy~\cite{bartsch2016control}. However, users find these privacy settings cumbersome~\cite{van2015social}, either preferring to stick to defaults~\cite{fiesler2017or} or setting them only once at the beginning~\cite{strater2008strategies}. Even when the user correctly configures the privacy of their posts, their connections can leak unwanted information. Collective nature of the online social media requires connections of the user to understand the implications of their interactions for privacy protection. Hence, in this work we aim to gauge the user understanding of public and protected account information and interactions visibility on Twitter along with user awareness of the reach of these posts. While most of these mentioned works are on platforms where privacy settings are granular, our work focuses on Twitter which has a relatively simpler privacy configuration.

There is limited work on user understanding of privacy settings and information visibility on Twitter. Proferes~\cite{proferes2017information} conducted a user survey with 434 participants to measure their understanding of Twitter in terms of techno-cultural and socioeconomic aspects. They provided participants various statements, in a range of topics including data, users, governance, algorithms, etc., and asked them whether the statement was accurate. They found that the participants did not understand the long term storage of past tweets as well as the visibility of information to other users. Only 24.7\% of the participants correctly answered that number of tweets, followers, followees etc. were public for protected accounts. While the statements covered some topics around account information visibility and interactions between accounts, they did not investigate which of those account information were understood better or what was the participants' expected audience for the interactions. 

In a study to understand the date of birth disclosures on Twitter, Kekulluoglu et al.~\cite{kekulluoglu2022authentication} asked their participants two interaction visibility questions between public and protected accounts. They found that participants commonly thought these interactions could happen but the audience will be limited to the followers of the protected accounts. Compared to these two studies, we expand on the account information questions and ask more detailed interaction questions in our work. We also divide our results depending on the account type of the users and compare their understanding.

\change{There are various factors that affect privacy knowledge such as frequency of internet access~\cite{park2014understanding}, frequency of SNS use, frequency of privacy settings usage~\cite{bartsch2016control}, and so on. Privacy literacy and privacy behaviour are also positively related~\cite{sindermann2021online}. Bartsch and Dienlin~\cite{bartsch2016control} surveyed 630 Facebook users and found that participants who changed their privacy settings more frequently had higher privacy literacy. Choi~\cite{choi2022privacy} found that people who use boundary regulation features (i.e. blocking) were more likely to have better privacy literacy. However, they also found that privacy rule application and privacy literacy were not significantly related. Proferes~\cite{proferes2017information} found that Twitter users were better informed about the functionality of the features they can see and interact, i.e. ``information flow solipsism'', which is also supported by Hagendorrf~\cite{hagendorff2018privacy}.} 

\section{Methodology}

We conducted a user survey to measure Twitter users' awareness of information and tweet visibility. We conducted a prescreening survey asking participants if their Twitter account is public, protected, or switches between the two. \change{Studies have shown that privacy literacy increases online privacy behaviours~\cite{sindermann2021online} and in turn usage of privacy settings can increase the privacy literacy~\cite{bartsch2016control}. Hence, it is expected that users who changed their privacy settings would have better understanding of the information visibility compared to public users.} We invited a balanced number of participants of each account type to take our survey so that we can measure the possible effect of account type on the user understanding. The main study consisted of questions around information and tweet visibility on Twitter along with demographic questions around their account. We recruited our participants from Prolific Academic (PA)~\cite{prolific}. 

The main survey was pilot tested with 6 PA participants before launch to estimate the time required to complete the survey and get feedback on the clarity of questions. We designed and ran both surveys following the University's ethics protocol and compensated each participant with £2.25 for filling the main survey (£9 per hour).

\paragraph{Prescreen} 
The visibility of a user's account likely impacts their understanding of Twitter's visibility settings~\change{\cite{bartsch2016control}} so we used a prescreen to ensure that we invited even numbers of people who are protected, public, and switch between. Doing so is especially important given that only 4\% of Twitter users have protected accounts~\cite{liu2014tweets} so prescreening is necessary to ensure adequate participation from all groups. 
We conducted the screening survey at the end of June 2021. We used PA and limited the survey visibility to those who can speak English fluently and have Twitter accounts using PA's filtering feature. We compensated each participant with £0.09 for filling the prescreen. 
We asked participants what their normal Twitter audience was on a 7-point Likert\change{-type} scale from: ``Always protected'' to ``Always public''. We then split respondents into three groups: 
\textbf{Public} (``Always public''), \textbf{Protected} (``Always protected''), and \textbf{Switching} (other answers). Out of 1074 users who had a Twitter account, 179 (16.7\%) were protected, 408 (38\%) were public, and the remaining 487 (45.3\%) switched. 
Following the prescreen, we invited equal number of participants from each group to take our main survey in July 2021.

\subsection{Survey Instrument}
After informed consent, we asked participants if they had a Twitter account (they all did) followed by four sets of questions around Twitter functionality with respect to different account types. The first set of five questions asked about the visibility of individual tweets posted by: public users, protected users, users who were protected but changed to public, users who were public but changed to protected, and lastly users who were public and stayed public.

Next, we asked questions about the visibility of 11 types of account information (e.g. followers, lists) for public and protected accounts.
Followed by 11 scenario-based questions about what would happen if a public and protected account attempted to interact in various ways (e.g. quote tweet, retweeting). Finally we asked four true/false statement questions targeting potential misconceptions involving how Twitter behaves towards different account types (e.g. users with protected accounts cannot be tagged in photos). 

The last section covered demographics including the normal audience question from the prescreening, frequency of Twitter usage, the information they have on their profile, number of followers, as well as the number of users they follow. We also asked if they can easily tell the type of account when replying and if they look at account type before engaging (reply, mention, retweet) with a tweet. Finally, we provided an optional free text comment box. Full survey text is available in supplementary materials. 

\subsection{Participants}
In total, 459 participants completed the survey, but 123 were excluded due to failing an attention check question resulting in 336 users. Common participant demographics (e.g. sex, age, nationality) as well as the demographics related to the filters set by researchers (e.g. social media usage) are provided by PA.   According to those demographics, 141 (42\%) of our participants were female and 193 (57.4\%) were male with 1 preferring not to respond. Our participants had an average age of 25.2 (sd 7.4) and a median of 23. 214 (63.7\%) of the participants were between 18-24 years old, 88 (26.2\%) were between 25-34 years old, 25 (7.4\%) were between 35-44 years old, and the remaining 8 (2.4\%) being 45 or older. Most of the participants used Twitter daily (194, 57.7\%) followed by weekly (95, 28.3\%), monthly (29, 8.6\%), and a couple times a year (18, 5.4\%). 

In regards to account type, 
75 (22.3\%) of the participants keep their accounts always protected while 131 (39\%) keep them always public, the remaining 130 (38.7\%) change their privacy settings. Out of those who change their privacy settings, 27 (8\%) stays mostly protected, 15 (4.5\%) somewhat protected, 14 (4.2\%) balanced, 20 (6\%) somewhat public, and 54 (16.1\%) mostly public. Twitter accounts are public by default and protected users need to change their settings only once, while users who switch indicate that they utilize the privacy settings more often than other users. For simplicity, we use the terms public users, protected users, and switching users in the remainder of the paper to refer to these account types. 

Most participants (62.5\%) had less than 100 followers. They themselves mostly followed between 100 and 499 users (50.9\%) or less than 100 users (37.8\%). 
82.4\% had a profile picture, 59.5\% had a header photo, and 70.5\% biographic information. In regards to more sensitive data, 33.6\% had their birthdays, 26.2\% their location, and 13.1\% had a website on their profiles. 

\paragraph{Interacting with Protected Accounts}
61.9\% of participants said they can easily tell when they are replying to a protected account (70.7\% of protected users, 63.4\% of public users, and 55.4\% of switching users). However, only 32.4\% said they check account type when they are engaging with a tweet (36\% of protected, 32\% of public, and 30.8\% of switching). 

We asked participants how often they interact with protected users via liking tweets, replying to their tweets, or mentioning the account.
36\% of our participants ``Always'' or ``Often'' like protected accounts' tweets, 17\% reply to them, and 8\% mention them.

\section{Results}

\subsection{User Awareness of Visibility \change{(RQ1)}}
We asked about the visibility of information, tweets, and Twitter functionality to our participants. 
Table~\ref{tab:totalResults} shows the percentage of correct answers given by our participants to each set of questions described in the survey instrument. In general, users were able to answer questions correctly. Over 80\% of questions about individual tweet visibility and account information visibility were answered correctly by our participants. But users had more trouble when answering questions that involved interactions between two accounts of different types. Misconception questions, which focused on less common Twitter actions, were the least understood with 39.1\% of the participants answered correctly. 

\begin{table*}[t]
	\centering
\begin{tabular}{p{6cm} c }
Question Set & Accuracy \\
\hline 
All Questions & 71.7\% \\
Individual Tweet Visibility & 86.7\% \\
Public Account Information Visibility & 85.9\% \\
Protected Account Information Visibility & 82.9\%\\
Interaction Visibility & 51.2\% \\
Misconceptions & 39.1\% \\
\hline
\end{tabular}
\caption{Percentages of correct answers given to each question set.}
\label{tab:totalResults}
\end{table*}

\paragraph{Individual Tweet Visibility}
We asked five questions about the visibility of tweets under different account types and in cases where the account tweets and then later changes type, the answer breakdowns are visible in Table~\ref{tab:individualVis}.
As a simple check, we started this section with two very easy questions asking who can see public account tweets and who can see protected account tweets. 87\% of the participants correctly answered that anyone on internet can see tweets posted by public accounts with the remaining 13\% incorrectly selected that only logged in Twitter users could see public account tweets, indicating an awareness of wide visibility but not properly understanding just how wide. For protected accounts, 95\% indicated that only the followers of a protected account can see the tweets. 
Answers to the other questions show less understanding of how historical tweets are handled when an account switches type. Though only 7\% incorrectly thought that when changing from protected to public, past tweets would stay protected and only 8\% thought that when changing from public to protected past tweets would stay widely visible to logged in Twitter users or anyone on the internet.

\begin{table*}[t]
	\centering
\begin{tabular}{p{6cm} c }
Individual Visibility Questions & Accuracy\\
\hline 
Who can see a public account's tweets & 86.9\% \\
Who can see a protected account's tweets & 95.5\% \\
Change visibility from public to protected & 89.3\% \\
Change visibility from protected to public & 76.2\% \\
Keep visibility setting public & 85.7\% \\
\hline
\end{tabular}
\caption{Percentages of correct answers given to tweet visibility questions for an account.}
\label{tab:individualVis}
\end{table*}

\begin{figure}
\centering
\begin{minipage}[t]{.485\textwidth}
\centering
  \includegraphics[width=1\linewidth]{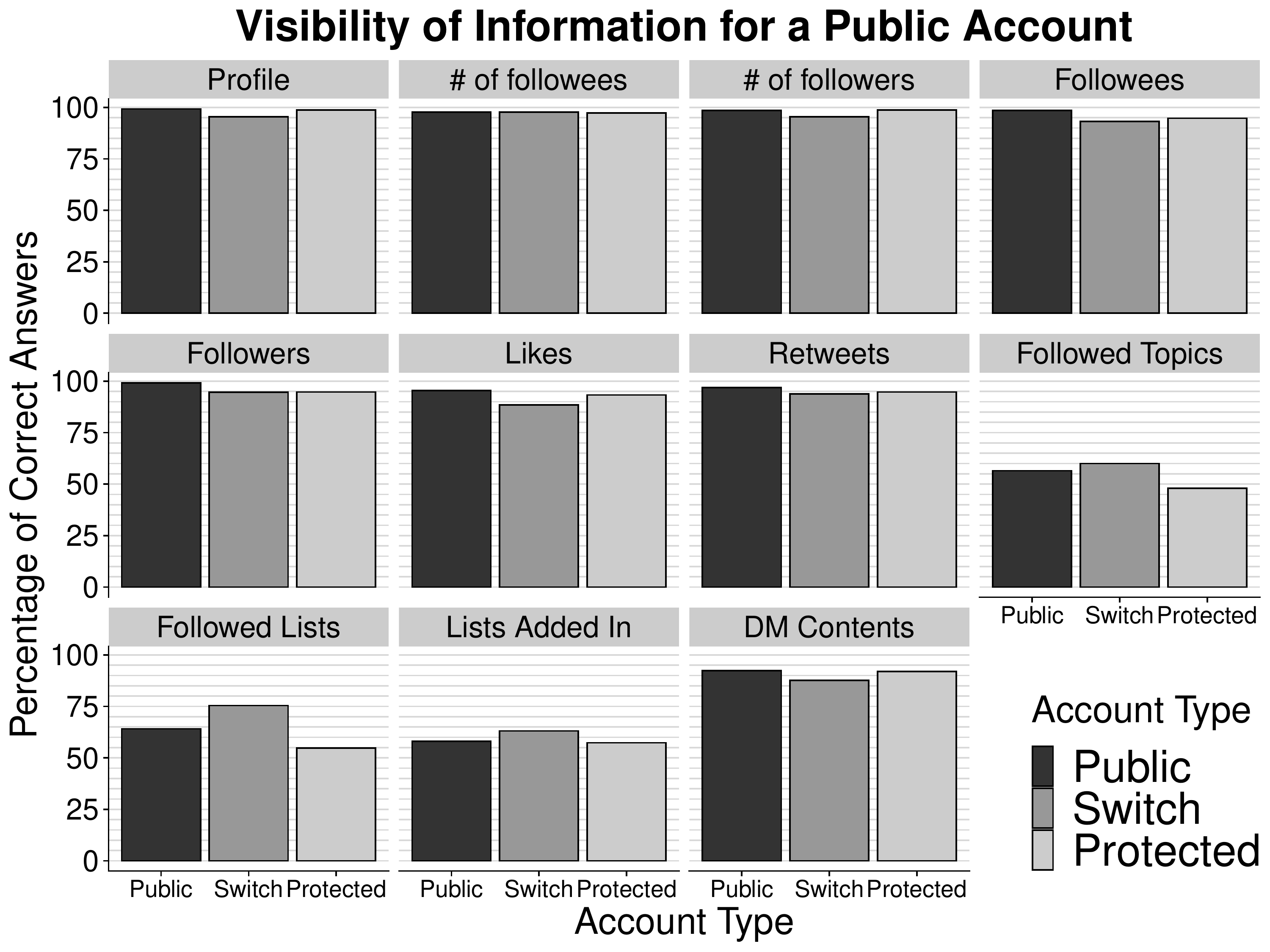}
  \vspace{-0.3cm}
   \caption{Percentage of correct answers given to account information visibility questions regarding a public account. }
   \label{fig:visInfoPub}
\end{minipage}%
\hfill
\begin{minipage}[t]{.485\textwidth}
\centering
  \includegraphics[width=1\linewidth]{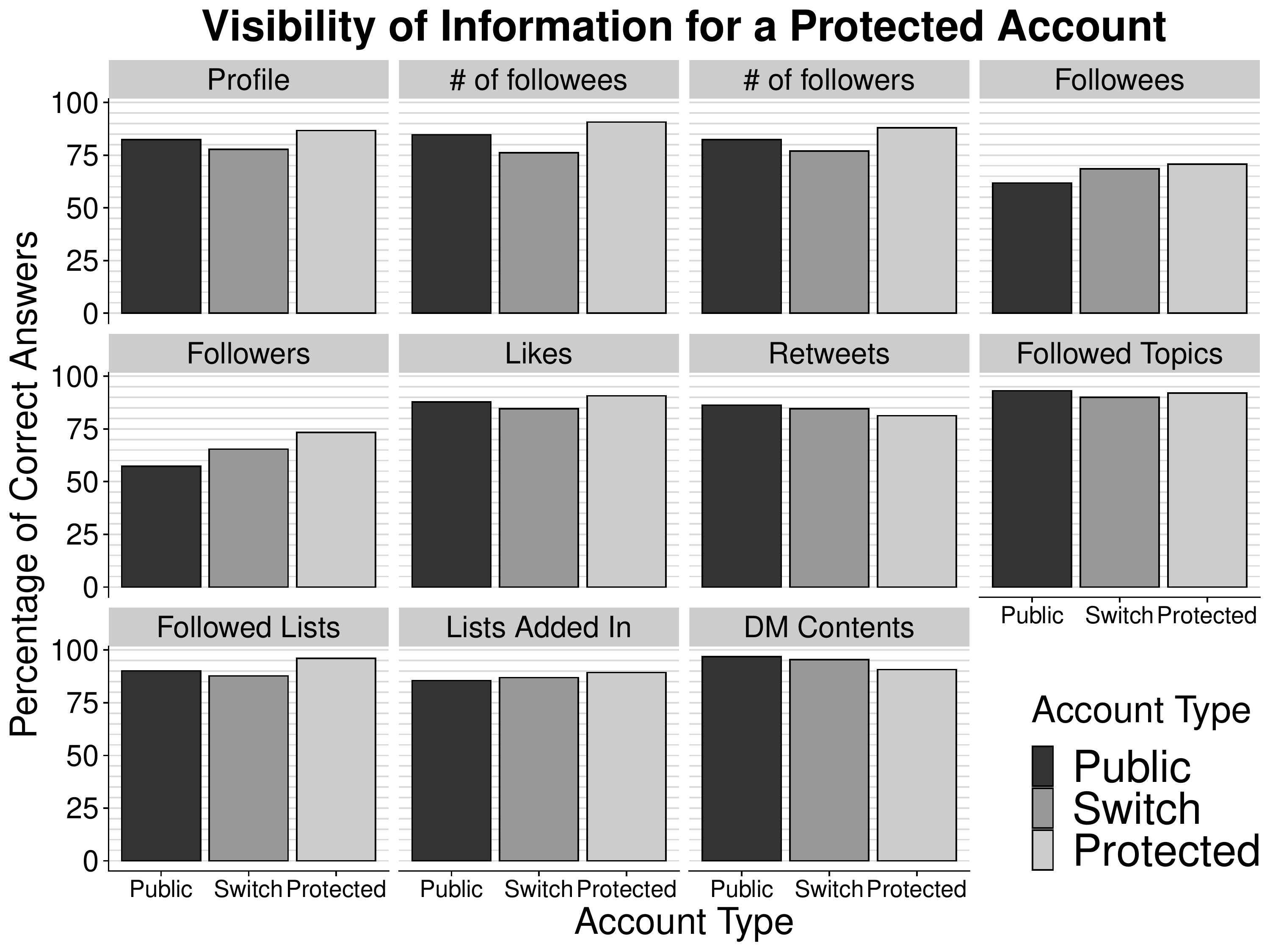}
  \vspace{-0.3cm}
  \caption{Percentage of correct answers given to account information visibility questions regarding a protected account.}
  \label{fig:visInfoPro}
\end{minipage}
\end{figure}

\paragraph{Account Information Visibility}
We asked about information visibility of public and protected account information to measure awareness of their visibility. We asked our participants about the visibility of the 11 types of information a Twitter account can have.   Figures~\ref{fig:visInfoPub} and \ref{fig:visInfoPro} show the percentages of correct answers given by participants. 
Currently, all of the information shown in Figure~\ref{fig:visInfoPub}, aside from the ``DM Contents'', are publicly visible for a public account. On the other hand, only ``Profile'', ``\# of followees'', and ``\# of followers'' are publicly visible for a protected account (Figure~\ref{fig:visInfoPro}). Most of our participants correctly selected whether the information was publicly visible or not for all types of information. However, the participants had lower understanding around topics and lists for public accounts. The visibility of followers and followees of protected accounts were also less understood by participants.

\begin{figure}
\captionsetup{justification=raggedright}  
\centering
  \includegraphics[width=0.75\linewidth]{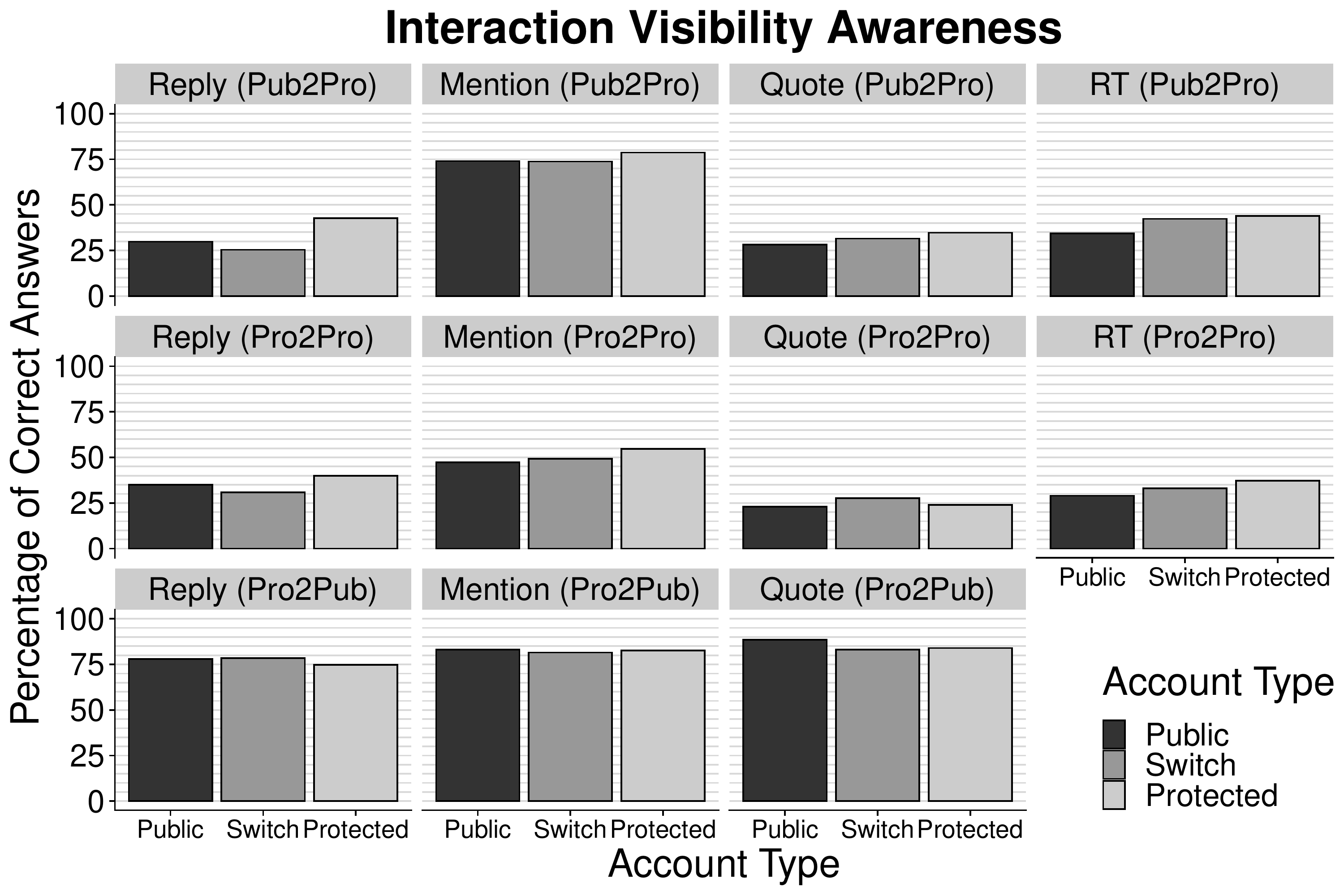}
  \vspace{-0.3cm}
   \caption{Percentages of correct answers given to interaction visibility questions. First row for interactions from public accounts to protected ones, second row for protected to protected interactions, and the last row for protected to public interactions.}
   \label{fig:interactionVis}
\end{figure}

\paragraph{Interaction Visibility}
We give the percentage of correct answers given by our participants to interaction visibility questions in Figure~\ref{fig:interactionVis}. We asked participants various scenarios between two accounts that follow each other and asked them what would happen if one of these accounts interacted with the other by replying, mentioning, quote tweeting, and retweeting. All of the questions included interactions with protected accounts.  The first question set, which had a public account interacting with a protected one, was challenging for participants with most participants answering incorrectly for replying, quote tweeting and retweeting. 
A common error involved a public account replying to a protected account's tweet, the public account's reply would be public, but 66\% of participants incorrectly thought that only the protected accounts followers could see the reply tweet. 

In the second scenario we asked what would happen if a protected account interacted with another protected account. For all of the interaction types, the majority of participants incorrectly answered that the interaction could happen but only users who follow both of the protected accounts could see it. Participants performed the best in the third scenario where we asked what would happen if a protected account interacted with a public one with over 75\% of participants answering all questions correctly. 

Based on their answers to these questions, participants clearly thought that when interacting with a protected account the tweet or other interaction would only be visible to the followers of the protected account. 
Since tweet visibility is only connected to the poster's account type, the possible interactions initiated by public accounts such as replying and mentioning will be visible publicly. If two protected accounts interact, then the followers of the account who initiated the interaction, i.e. reply and mention, are the only ones who can see those tweets.

\paragraph{Misconceptions Around Account Type and Twitter Functionality}
In addition to the visibility questions we also asked four questions around Twitter functionality with different account types \change{in Table~\ref{tab:misconceptions}.}Our participants had the lowest performance in this set of questions which somewhat expected for this question group. 
Direct message (DM) requests and picture tagging are controlled by a different set of settings than the public/protected account type setting which is potentially confusing.
Which turned out to be the case, less than half the participants knew that protecting an account will not disable tagging that account in pictures. 
Worse, only 18.8\% of participants knew that protecting an account does not disallow non-followers from sending DMs to that account.

Lastly, we asked two questions around tweet deletion and the residual data~\cite{mondal2016forgetting}. Replies to a deleted tweet stay in the platform even when the deleted tweet was posted by a protected account. Most of our participants (55.1\%) correctly said that deleting a public account's tweet will not delete replies to it. However, only 36\% correctly answered when the same question was asked regarding a protected account.

\begin{table*}[t]
	\centering
\begin{tabular}{p{7cm} c | c c c} 
Misconceptions around functionality & Total & Public & Switching & Protected\\
\hline 
Protected accounts cannot be tagged in photos & 46.4\% &  48.1\% &  47.7\% & 41.3\% \\
Protecting an account will disable DMs from non-followers & 18.8\% & 10.7\% & 23.1\% &  25.3\% \\
Deleting a public account's tweet will delete replies to it & 55.1\% &  53.4\% & 56.9\% & 54.7\% \\
Deleting a protected account's tweet will delete replies to it & 36\% &  35.1\% & 35.4\% & 38.7\% \\
\hline
\end{tabular}
\caption{Percentage of correct answers given to questions around Twitter functionality for different account types. Answers divided based on the account type. Percentages are out of the total number of public, switching, and protected participants respectively.}
\label{tab:misconceptions}
\end{table*}

\subsection{Factors That Contribute to User Awareness \change{(RQ2)}}
We now investigate to what extent user characteristics affect their ability to determine who can see their information. For this purpose, we construct a generalised linear model that contains the main factors which explain variation in the accuracy of answers. 

\change{When considering a small number of predictors, typically, all of these can be entered into the model at the same time. However, the design of this study is more complex. We hypothesise that four groups of predictors might affect performance;

\begin{description}
\item[Social Media:]  Social media experience on platforms other than Twitter.
\item[Profile Elements:] Amount and type of  information included in the Twitter user's profile that might make them identifiable  
\item[Audience:] Typical audience of the user.
\item[Twitter-specific:] Aspects of the user's behaviour on Twitter except for audience
\end{description}

Entering all predictors into a single model is likely to lead to a model with low generalisability. Therefore, we perform a two-stage predictor selection process. In the first stage, for each group of predictors, we select those that explain the highest amount of variation in participants' performance. This results in the four models summarised in Table~\ref{tab:aicmodels}. 
In the second stage, we use the same selection strategy to construct a combined model from the predictors used in the four group-specific models. The resulting model combines aspects of all four predictor groups. The predictors and their relative importance in accounting for variation in participant performance are shown in Table~\ref{tab:fullmodel}. }

\paragraph{Method} 

Generalised linear models are an extension of linear regression models. Here, we use a type of generalised linear model that is adapted to binary outcome variables, namely logistic regression. 
The outcome variable we are looking at is \emph{accuracy} - whether the respondent answered a question about tweet or account information visibility correctly.

We use the term ``model'' since the aim of model building is to construct a generalised linear model, described by a logistic regression equation, that explains the highest amount of variation in the data set in the most parsimonious way possible. Thus, we assess model quality using the well established AIC (Akaike Information Criterion), which rewards models that fit the data well, and penalises models with many parameters. When comparing multiple models of the same data set, a lower AIC is better.

We have already seen that there are some aspects of tweet visibility that users understand well, and others that they struggle with. Therefore, our baseline model is $accuracy = 1 + question$, where $question$ stands for one of the aspect of Twitter functionality and tweet visibility that are covered in the main body of the questionnaire.

We then use a process of greedy stepwise selection to add the most appropriate $m$ user factors to the model. These factors are selected from a fixed set of $n$ candidates. The procedure used is as implemented in \texttt{stepAIC} in R, MASS package. \change{Categorical variables that represent candidate factors are entered into the model using standard dummy coding~\cite{fox:02}}

First, we construct four models that cover \change{the different aspects of users highlighted above} and compare their performance. The models are:
\begin{description}
\item[Social Media:] whether the user has social media experience on Facebook, Tiktok, Instagram, Snapchat, or Linkedin
\item[Profile Elements:] the types of information the user includes in their Twitter profile (bio, website, picture, header, date of birth, location)
\item[Twitter-specific:] This includes all twitter-specific features except for audience, such as number of followers (ordinal scale converted to numeric), frequency of Twitter use, whether they can easily tell the type of account when replying, whether they look at the account type before engaging, and whether they like, reply to, or mention protected accounts.
\item[Audience:] The typical audience of the user. We use two variables, audience (always public, always protected, switching between public and protected), and the original seven item Likert\change{-type} scale (audience7).
\end{description}

We then compared how well each of the resulting four models explain the amount of variation in the data set using Anova, with the Chi Square test to establish significant differences. Finally, we used greedy stepwise selection to construct a final, full model using all features from the four individual models.

\paragraph{Results}

\begin{table}[]
    \centering
    \begin{tabular}{p{2.5cm}|p{9cm}|r}
    \hline
    Model & User-Specific Variables Included In Order & AIC \\ \hline
        Audience & Audience7 & 12829 \\
        Profile Elements & header, profile picture, website, location, date of birth & 12784 \\
        Social Media & on Instagram, on LinkedIn, on Tiktok, on SnapChat, on Facebook & 12776\\
        Twitter-specific & can easily spot protected accounts, replies to protected accounts, usage frequency (numeric), number of followers & \textbf{12607} \\ \hline
    \end{tabular}
    \caption{Performance of Four Models Covering Different Aspects of Users. AIC = Akaike Information Criterion. Variables listed in the order in which they were included in the model during stepwise greedy selection. All models also include an intercept and the variable for Question, which is part of the baseline.}
    \label{tab:aicmodels}
\end{table}

Table~\ref{tab:aicmodels} summarises the four models created using different user characteristics. In the model specifications, variables are listed in the order in which they were added to the model in the stepwise selection process.

We see clearly that the model which only takes audience into account performs worst. Use of other social media platforms yields a somewhat better model, just as considering the information disclosed in a person's Twitter biography. However, the model that explains the data best is the one that focuses on people's knowledge of Twitter, the number of followers, and frequency of use. When comparing the amount of variation explained by the four models using analysis of deviance (\texttt{anova} function in R), the Twitter-specific model outperforms all others with $p<0.00001$, and the audience-only models is worse than the model based on profile elements alone  ($p<0.00001$) and the model based on other social media activity alone ($p<0.00001$).

In other words, the initial four models show that users who use the ``protected'' switch do not necessarily understand what it does---users who frequently engage with Twitter, know how to spot protected accounts, and have a sizeable number of followers do. Participants who strongly agreed that they can easily tell the account they are replying to is protected answer 76.7\% of  all questions accurately, while participants who answered strongly disagree perform worse with 64.5\%. Participants who reported they always reply to protected accounts get 80.7\% of questions right, where this rate is 69.3\% for participants who never replies to protected accounts' tweets. Finally, participants who used Twitter daily answer 73.5\% of questions correctly, while participants who reported they use Twitter a couple of times a year perform at 62.7\%.

The full model, as selected from all variables in the four previous models, is shown in Table~\ref{tab:fullmodel}. We do not show the complete logistic regression model with all coefficients since there are more than 60. Instead, we focus on the relative importance of the variables included in explaining the variation in the data set. In Table~\ref{tab:fullmodel}, variables are listed in the order in which they were added to the model in the greedy stepwise selection process. The p-value given for each variable indicates whether adding the variable yields a significant improvement over the previous model that did not include the variable. For example, once the model includes the first seven variables in Table~\ref{tab:fullmodel}, up until the number of followers, the next best addition is information about whether the user is on Snapchat ($p<0.05$). Once information on Snapchat use has been integrated, the next most important information is whether the user is on Tiktok ($p<0.005$)

We see that, again, the most important variables indicate whether users can easily spot protected accounts, and whether they engage with protected accounts. Experience with Instagram and LinkedIn is also important. While the type of information mentioned in the user's profile and the audience setting (public, protected, different levels of switching) do cover significant additional variation in the data set, their relative contribution, as indicated by the reduction in residual deviance, is small. 

\begin{table}[]
    \centering
    \begin{tabular}{l|rrr}
\hline
Variable & df & Deviance & Pr(>Chi) \\ \hline

Question &           41   & 4074.4   &   $p<0.00001$ \\
Can easily spot protected accounts       &  4  &  174.0  &    $p<0.00001$ \\
Replies to protected accounts    &  4    &    42.2    &    $p<0.00001$ \\
User on Instagram      &   1    &    24.0    &    $p<0.00001$\\
Frequency of use      &  1    &    17.0    &    $p<0.00001$ \\
User on LinkedIn      &  1    &    13.8    & $p<0.0005$ \\
Number of Followers   &     4   &     19.8   &  $p<0.001$  \\
User on Snapchat     &   1     &    4.8  &   $p<0.05$    \\ 
User on Tiktok       &  1   &      9.1   &    $p<0.005$  \\
Has profile picture     &  1    &     3.8    &     $p<0.1$ \\  
Audience            &  1   &      4.7   &   $p<0.05$    \\ 
Web site in profile & 1    &     4.2   &  $p<0.05$      \\ 
Location in profile & 1   &   4.3    &   $p<0.05$  \\ \hline
    \end{tabular}
    \caption{Relative Importance of Each Factor in the Order in Which It was Added to the Model. Df: degrees of freedom. Deviance: measure of variation in the data set covered by variable.  Pr(>Chi): probability that the model with variable $x_{i}$ is an improvement over the model with variables $x_1,\ldots,x_{i-1}$}
    \label{tab:fullmodel}
\end{table}

\section{Discussion}

One of the important aspects of protecting privacy in online social networks is understanding who can see the information and the posts that are being shared. Previous work has shown that users struggle to configure privacy settings to reflect their privacy expectations~\cite{madejski2011failure,liu2011analyzing} and they are confused with different privacy settings~\cite{sleeper2013post}. However, these studies are done on social media platforms that have more granular privacy options than Twitter. The findings in our study show that users confusion with privacy settings occur even with the relatively simplistic binary Twitter privacy settings. These results are in line with Proferes'~\cite{proferes2017information} work on user beliefs about Twitter, though in some cases our participants did show better understanding. In their study, only 24.7\% of the participants correctly answered the number of tweets, followers, followees, and so on that would be public for protected accounts. However, 80\% of our participants correctly answered for both the number of followers and followees. Considering the seven years between the two studies, user awareness of such settings could have changed. Its also possible that since our sample includes more protected and switching users than a random sample, that is impacting the level of awareness, though given our own results, the impact is likely minimal.

Tweets from public accounts can be seen by anyone on internet while tweets from protected accounts can only be seen by approved followers. Individual tweet visibility is well understood by our participants with over 85\% of questions correctly answered. However, who can see the tweets when these accounts interact is not necessarily as clear. In our study, most of the participants could not answer what would happen if an account interacted with a protected one (excluding mentioning). Less than 40\% knew that protected account tweets could not be retweeted, and only 31\% were aware that anyone could see the reply by a public account to a protected account's tweet. \change{The knowledge of participants who have a public, switching, or protected accounts was not significantly different which suggests that the user's account type alone does not improve or worsen their knowledge of how Twitter privacy works. This is surprising since we expect better privacy awareness from protected users and the switching users since they limit their information visibility to protect privacy~\cite{debatin2009facebook}. Especially the switching users, who interact with the privacy settings more frequently and change their settings back and forth to regulate their boundaries~\cite{kekulluoglu2022understanding}, should have better knowledge of the functionality~\cite{bartsch2016control}. According to }Bartsch and Dienlin~\cite{bartsch2016control}, the time spent on Facebook and the frequency of utilizing the privacy settings on the platform led to better online privacy literacy. Similarly, we find that frequency of Twitter usage is a factor that contributes to the knowledge of the information and tweet visibility. However, \change{as mentioned} the participants' normal audience did not have any significant impact on the awareness. \change{This is contrary to prior work that show people who use privacy settings have better privacy literacy~\cite{sindermann2021online,bartsch2016control}. Our results were in line with Choi~\cite{choi2022privacy} who found that the relation between privacy literacy and privacy rule application were insignificant.}


 %

\change{Privacy protection on online social network platforms is not an individual task~\cite{altman1976conceptual,petronio2015communication}. Interactions with a post can disclose information about the owner, even when the post is hidden~\cite{kekulluoglu2020analysing}. Communication Privacy Management (CPM) Theory~\cite{petronio2015communication} defines ``boundary turbulence'' when there is a conflict between the privacy rules of the information owner and the co-owners, e.g. the people the post is shared with. As a result of this conflict, the information is shared with unintended third parties. People may avoid sharing information and interacting with their connections because of their privacy concerns~\cite{sleeper2013post} which can be harmful, especially for vulnerable populations~\cite{dym2018vulnerable} and people who seek support by sharing experiences~\cite{andalibi2018announcing}. This work shows that some users are not even aware of the caused turbulence and assume that the visibility of their interactions are limited by the privacy rules of the more restrictive party in the interaction.}

\change{There are some steps users can take to understand the functionalities of the platforms and the implications of their interactions. It is important to get familiar with the privacy settings of a newly signed-up platforms. It is common for users of social media platforms to keep the default settings provided~\cite{fiesler2017or} but they are also mostly unsure what these default settings entail~\cite{proferes2017information}. Even if a person does not want to change their settings, observing the interface can increase their awareness about their interactions with users who have different settings. Another step users could take is to check the insights provided by the platforms. For example, Twitter gives analytics about the tweets where users could look at the number of impressions and engagements of their posts. Since users tend to underestimate the size of their audience~\cite{bernstein2013quantifying}, checking the insights could help to reduce the gap between the imagined audience and the real audience.}

While our sampling strategy resulted in 22\% of our participants having always protected accounts and a further 40\% sometimes having protected accounts, the general Twitter population has a much lower percentage of protected accounts \change{(4\%)}~\cite{liu2014tweets} with only  13\% of the US adults choosing to have their Twitter account protected~\cite{wojcik2019sizing,pewprotected}. Given that, it is somewhat surprising that 81\% of the Always Public participants indicated they had liked, mentioned, or replied to a protected account. Crowdworkers on websites like Mechanical Turk are known to have more privacy concerns than average users~\cite{mturk-privacy} which may explain some of the finding. But the finding still suggests that despite being uncommon, public accounts do interact with protected ones, more than might be expected based purely on the number of protected accounts. 
Surprisingly, protected account holders answered the interaction questions in a similar way to public account holders, suggesting that even though they actively chose to have their tweets private with their followers only, they share similar misconceptions and may not be aware that half of their conversations with public accounts are visible to everyone, possibly violating their intended privacy outcomes. 

Compared to the general public, Twitter users are younger and more educated~\cite{wojcik2019sizing} as well as being more skilled in using internet~\cite{van2011rethinking}. Having better internet skills is shown to have a strong effect on privacy protection~\cite{buchi2017caring}. In addition, the 42\% of our survey participants were female, who are shown to be more privacy conscious~\cite{fiesler2017or}, where only 31.9\% of all Twitter users are female~\cite{twitter-gender}. \minorchange{This difference in representation can be a result of our recruitment strategy since we balance the account types and female social media users are more likely to limit their visibility~\cite{fiesler2017or}. }Despite these factors our participants had a low understanding of the visibility of information around topics and lists, as well as the visibility of tweets when interacting with protected accounts. A recent article also states that Twitter is aware of the low understanding their users have of privacy settings~\cite{twitter-privacy-features}.

\subsection{Design Implications}

Our findings suggest that the two main points of confusion about information visibility on Twitter are: 1) interactions, especially with protected users and 2) topics and lists. We discuss the privacy implications of these points and give design suggestions for helping user understanding below.

\paragraph{Showing potential audience when drafting tweets:} Users have an imagined audience when they share posts~\cite{vitak2015balancing}, which is often smaller than the real audience size~\cite{bernstein2013quantifying}. Underestimating the audience of the tweets can lead to privacy problems, especially when interacting with protected accounts~\cite{kekulluoglu2020analysing}. Showing users the potential audience of their tweets can help users to better contextualize the reach of their tweets~\cite{lieberman07,vaniea2012}. For example, when users with public accounts are interacting with protected accounts, it could be stated that the tweets can be seen by everyone, not only the followers of that protected account. 

\paragraph{Hiding interactions with protected accounts:} Twitter UI prevents retweets/quote tweets of protected tweets, hence the effects of not being informed on the expected behavior is minimal. However, our findings suggest that big chunk of users believe that their tweets can only be seen by the followers of the protected account when they engage with one. This may lead them to disclose information they did not intend to share with the general public. Searching a person's account (e.g. ``@username'') in Twitter will bring up the tweets sent to them such as mentions and replies, even when they are a protected account. Hiding these tweets from non-followers can help protect the privacy of the protected accounts, which is closer to what users tend to believe currently~\cite{kekulluoglu2022authentication}.

\paragraph{Showing types of the interacted accounts clearly:} Even if the users were perfectly aware who could see their tweets when they interact with protected accounts, it is possible that these users are not aware they are interacting with one. Only 29\% of our participants strongly agreed that they can easily tell that they are replying to a protected account, while this rate was 10\% for checking the account types of the users they engage with. Our analysis showed that the participants who were able to easily tell that they are replying to a protected account had higher awareness of the platform. Its unclear the direction of the relationship though. It could be that those who are more conscious of protected users' privacy are consequently more self-assured in their ability to notice such accounts, or it could be that the ability to notice such accounts causes users to become aware.  Design changes to the interface could make protected accounts more visible, especially in cases where users interact with multiple accounts (i.e. replying to a reply).

\paragraph{Informing users about the visibility of topics/lists:} Our participants were mostly aware whether different types of account information are publicly visible. The visibility of lists and topics on public lists was confusing, both are publicly visible by anyone for public accounts along with the lists a user has been added to. Users can follow topics they are interested in and create lists to curate a timeline of users they want to keep track of without necessarily following individual accounts. While there is an option to create private lists, there is no option to hide the topics that a public account follows on Twitter. Public users cannot hide the lists they are following which are created by other users. They also cannot remove themselves from a list without blocking the list creator. Blocking a user may create discomfort and users might not be willing to block another user especially if that person is a close friend or a family member~\cite{rashidi2020s}. These public topics and lists can leak information about the user including their interests and personal ties. Users should be clearly notified that this information around topics and lists is public. 

\paragraph{\change{General Suggestions for Social Media Platforms:}} \change{The design implications we provided are relatively easy to implement prescriptions~\cite{sas2014generating} that are specific to Twitter. However, these suggestions can be generalized to social media platforms. Social media platforms should state the potential audiences of the posts while they are being created. This is especially important when there is a ``boundary turbulance''~\cite{petronio2015communication} where there is a difference in privacy rules between different stakeholders (e.g. poster, mentioned people, and so on) of the post. The interface for interactions should also clearly lay out the stakeholders of the posts for posters to act accordingly. Functionalities and privacy implications of features, especially newly introduced ones, should be relayed to users in an effective way instead of being buried in help pages.}

\subsection{Limitations}
Many of the more specific limitations of the study are already presented in the discussion in relation to their findings. More generally, our study recruited from Prolific Academic which can draw participants from more privacy-conscious crowd, as shown for Amazon Mechanical Turk~\cite{mturk-privacy}. In addition, our sample is younger than global Twitter users~\cite{twitter-age} which also may translate into having better internet and privacy protection skills~\cite{van2011rethinking,buchi2017caring}. \change{Our sample also had a higher share of female participants than Twitter normally has~\cite{twitter-gender}. }This actually might show that the general public might be less aware of the visibility of their information and tweets. The study also uses a survey approach which means we are able to ask about only issues and answer options we know about in advance. We countered this issue by familiarizing ourselves with Twitter's range of options and ensuring that the full range was presented to users. We also endeavored to be comprehensive and clear in our question and answer presentation.  Finally, we included a comment box at the end of the study in case participants noticed anything they strongly felt was missing, reviewing these comments resulted in no serious identified omission.

\change{We did not measure whether our participants used the official Twitter apps or third-party client ones to access Twitter. Third-party client apps have different interfaces that could affect the user understanding. However, we decided to discuss the visibility and interface considering Twitter's own apps since only around 1\% Twitter users have Twitter clients installed~\cite{twitterclients}.}



\section{Conclusion}
We conducted a user survey with 336 participants to understand users' awareness of the visibility of information and tweets shared on Twitter, including the tweet visibility when accounts with different types interact. Our findings suggest that users are mostly aware information shared on accounts depending on the account type. They also understand the the audience of tweets sent by different account types when those tweets are not interacting with others by mentioning or replying. However, the functionality and visibility of interactions between accounts are not clear, especially when the interacted account is protected. Our participants tend to think these interactions will only be shown to the followers of the interacted protected account. Surprisingly, being a protected or switching account holder did not translate into a better understanding of interactions with protected accounts. Frequently using Twitter and being able to easily tell that they are replying to a protected account contributed to better performance in our participants. Informing users on how engagements work between different accounts is essential. Users also should be notified better when they are interacting with protected accounts.

\begin{acks}
\minorchange{We thank Nadin Kökciyan for the valuable feedback on the survey. We thank everyone associated with the TULiPS Lab and SMASH Group at the University of Edinburgh for helpful discussions and feedback. This work was supported in part by the EPSRC DTA award, funded by the UK Engineering and Physical Sciences Research Council and the University of Edinburgh.}
\end{acks}

\bibliographystyle{ACM-Reference-Format}
\bibliography{references}
\end{document}